\documentclass[aps,prb,twocolumn,groupedaddress,amsmath,amssymb]{revtex4}

\usepackage{amssymb}
\usepackage{amsmath}
\usepackage{graphicx}
\usepackage{subfigure}
\usepackage{textcomp}
\usepackage{color}
\usepackage{amsfonts}
\usepackage{bbold}
\usepackage{dsfont}
\usepackage{epsfig}
\usepackage{hyperref}

\begin{document}

\title{Structural relaxation effects on interface and transport properties of Fe/MgO(001) tunnel junctions}

\author{
Xiaobing Feng, O. Bengone, and M. Alouani\\
Institut de  Physique et Chemie des  Mat\'eriaux de Strasbourg, UMR 7504 CNRS-ULP, 23 rue du Loess,
Strasbourg 67034, France\\
S. Leb\`egue \\
Laboratoire de Cristallographie et de Mod\'elisation des Mat\'eriaux Min\'eraux 
et Biologiques UMR UHP-CNRS 7036, Nancy, France \\
I. Rungger and S. Sanvito\\
School of Physics and CRANN, Trinity College, Dublin 2, Ireland }

\begin{abstract}
The interface structure of Fe/MgO(100) magnetic tunnel junctions predicted by
density functional theory (DFT) depends significantly on the choice of exchange
and correlation functional. Bader analysis reveals that structures obtained by
relaxing the cell with the local spin-density approximation (LSDA) display a
different charge transfer than those relaxed with the generalized gradient
approximation (GGA). As a consequence, the electronic transport is found to be
extremely sensitive to the interface structure. In particular, the conductance
for the LSDA-relaxed geometry is about one order of magnitude smaller than that
of the GGA-relaxed one. The high sensitivity of the electronic current to the
details of the interface might explain the discrepancy between the experimental
and calculated values of magnetoresistance.
\end{abstract}
\pacs{71.15.Ap, 71.20.-b, 71.20.Eh}
\maketitle

Tunnel magneto-resistance (TMR) is the change of the electric resistance of a
magnetic trilayer made of two ferromagnetic metallic electrodes separated by an
insulating spacer, when the mutual orientation of the electrodes'
magnetizations changes from parallel alignment (PA) to antiparallel alignment
(AA). The figure of merit for the effect is the TMR ratio, defined as
$$ \mathrm{TMR}=\frac{G_\mathrm{PA} - G_\mathrm{AA}}{G_\mathrm{AA}}\:,$$
where $G_\mathrm{PA}$ ($G_\mathrm{AA}$) is the conductance for the PA (AA)
configuration.  Among the possible material combinations Fe/MgO(100) magnetic
tunnel junctions (MTJs) have attracted much attention, since extremely large
TMR ratios were theoretically predicted on the basis of symmetry-driven spin
filtering from the tunnel barrier\cite{Butler01}. These have been now
demonstrated experimentally\cite{bowen01,Yuasa04,halley07} and the
symmetry filtering effect is now widely accepted. However, the
measured TMR ratios are usually significantly lower than those
predicted by \textit{ab initio} methods. 

The origin of such a discrepancy is currently a matter of debate. On the theory
side, interface resonance states (IRSs) located around the Fermi energy
($E_\mathrm{F}$) are important for the zero-bias transport \cite{Ivan} and are
usually difficult to describe accurately. Moreover the MgO band-gap is
significantly underestimated by LSDA and GGA, so that the barrier height might
not be accurately predicted. On the experimental side, the quality of Fe/MgO
MTJs depends on the preparation methods. The 4\% lattice mismatch between Fe
and MgO produces dislocations and significant relaxation at the interface. In
addition, in typical growth conditions the Fe layers close to the interface
might be partially oxidized.\cite{schneider1}

Several different geometries for the Fe/MgO interface have been proposed
theoretically, based on relaxing [Fe]$_n$/[MgO]$_m$ supercells, consisting of
$n$ Fe and $m$ MgO monolayers (MLs) subject to various constraints. In
Ref.~[\onlinecite{Butler01}] LSDA relaxation for a [Fe]$_5$/[MgO]$_5$
superlattice, in which the in-plane lattice constant is held at the LSDA value
for bulk Fe, yields a Fe-O distance of 2.17 \AA. In contrast an early LSDA
calculation for one Fe ML on MgO substrate predicted an Fe-O distance of 2.3
\AA.\cite{Li91} Likewise a GGA study for one MgO ML on a Fe slab returns a Fe-O
distance of 2.21 \AA, and a separation between the first and the second Fe ML
6\% smaller than that of bulk Fe.\cite{Yu06} How relevant calculations for
single MLs on surfaces are for Fe/MgO MTJ is however not clear and there is
still disagreement between experiments \cite{Urano88} and theory \cite{Li91}.
These controversies over the correct interface structure bring the question of
how the different interface geometries affect the electronic and transport
properties. Our Letter aims at answering this question. 


%
\begin{table*}[t] \begin{center} \begin{tabular}{ccccccccc} \hline\hline DFT &
	$d_{-4}$ & $d_{-3}$ & $d_{-2}$ & $d_{-1}$ & $d_{0}$ & $d_{1}$ & $d_{2}$
	& $d_{3}$ \\\hline Unrelaxed&1.433&1.433&1.433&1.433& 2.160
	&2.026&2.026&2.026\\ 
LSDA&1.297&1.313&1.343&1.120& 2.002 &2.130&2.119&2.119\\ 
GGA&1.380&1.414&1.427&1.350& 2.219 &2.199&2.177&2.185\\
\hline\hline \end{tabular} \end{center} 
\caption{(Color online) The interface
structure of a [Fe]$_{10}$/[MgO]$_6$ superlattice obtained by LSDA and GGA
relaxation. The structural parameters for the unrelaxed structure are taken
from Ref.~[\onlinecite{Butler01}]. The middle two Fe layers are frozen at a
separation of 1.43 \AA\ and the experimental lattice constant for bulk Fe is
used for the in-plane lattice constant.  $d_n$ indicates the separation (in
\AA) between layers from Fe/MgO the surface ($n=0$), with the index $n>0$
($n<0$) labels MgO (Fe) layers.} \label{table1} \end{table*}
We perform structural relaxations for a [Fe]$_{10}$/[MgO]$_6$ superlattice with
both the LSDA and GGA\cite{PW91} functionals by means of the PAW
method\cite{Blochl94} as implemented in the VASP code\cite{Kresse93}. Charge
transfer and magnetic moments are calculated for the differently relaxed
structures with Bader analysis\cite{Bader65} using the LSDA densities.  A $12
\times 12$ in-plane ${\mathbf k}$-point grid and an energy cutoff of 400 eV are
employed to converge the charge density. As benchmark we find for bulk Fe a
lattice constant of 2.75~\AA\ and a bulk modulus of 2.68 Mbar, both in good
agreement with other all-electron LSDA calculations.\cite{Moroni97} Then, the
transport
properties are computed with the {\sc smeagol}\cite{Rocha06} code, which
combines the non-equilibrium Green's functions (NEGF) formalism with
DFT.\cite{siesta1} The GGA PBE\cite{PBE} functional and a $400 \times 400$
${\mathbf k}$-point grid are used for calculating the transmission coefficient. 

The layer spacings at the Fe/MgO interfaces obtained after structural
relaxation are summarized in Tab.~I.  These have been obtained by keeping fixed
the in-plane lattice constant, set equal to the experimental value for Fe, and
the atomic position of the two most internal Fe MLs. In the table we report the
obtained layer spacings along the stack direction $d_n$, where $d_0$ represents
the Fe-O distance at Fe/MgO interface and $n>0$ ($n<0$) refers to separation
between MgO (Fe) MLs. In the table we also report the unrelaxed structural
parameters taken from reference~\cite{Butler01}.

LSDA relaxation gives a Fe-O distance of 2.0 \AA. This is significantly
different from 2.16 \AA~of the unrelaxed structure. The LSDA relaxation also
predicts that the spacing, $d_{-1}$, between the first two Fe layers next to
the Fe/MgO interface is drastically reduced to 19\% of that of bulk Fe. The
changes in layer spacing of the MgO barrier on the other hand are relatively
small. Interestingly, the GGA results are significantly different from the LSDA
ones. In particular, GGA returns a Fe-O distance of 2.22 \AA\ and the reduction
of $d_{-1}$ is less pronounced than that for the LSDA. Our GGA-relaxed
structure is in good agreement with other GGA studies\cite{Yu06,bluegel1} and
rather close to the reference unrelaxed interface structure.

\begin{figure}[h]
\includegraphics[width=7cm, clip=true]{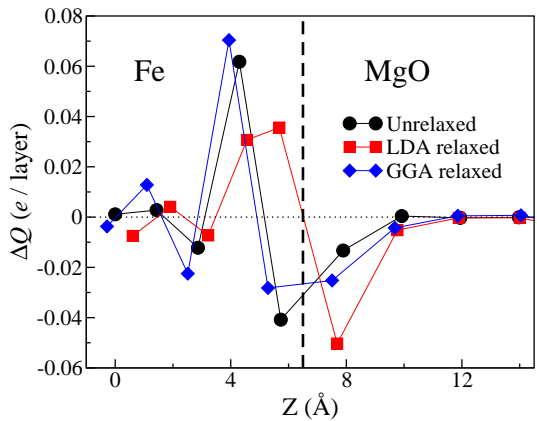}
\caption{(Color online) Charge transfer, $\Delta{\cal Q}$, for a [Fe]$_{10}$/[MgO]$_6$ superlattice as calculated from
Bader analysis. Net charge transfer is obtained by subtracting the atomic charges of bulk Fe and bulk MgO from the 
Bader atomic charges in the Fe/MgO superlattice. The vertical dashed line denotes the Fe/MgO interface.}
\label{charge_transfer}
\end{figure}

Fig.~\ref{charge_transfer} demonstrates that the charge transfer between Fe and
MgO is very sensitive to the interface structure. The Bader analysis for the
LSDA-relaxed geometry predicts that the first two Fe MLs at the Fe/MgO
interface acquire a similar amount of electron charge, $\Delta{\cal Q}$, while
for the GGA-relaxed and unrelaxed structures it indicates that the interface Fe
ML loses electrons and the second Fe ML gains a significant amount of charge.
In all cases MgO loses electrons to Fe. Similar charge transfer from MgO to
neighboring metallic layers was predicted before for Rh/MgO
interfaces.\cite{Castleton08} For the LSDA-relaxed structure we predict a net
electron transfer from MgO to Fe of up to about 0.06 electrons per atom.  This
is larger than the one for the other two geometries. The Bader charges for O
atoms at the interface and in the middle of the MgO barrier are respectively
7.64 and 7.71. This indicates that charge is transferred to Fe mainly from O.
Finally, since the GGA-relaxed structure is close to the unrelaxed one, the
charge transfer is also similar.

\begin{figure}[ht]
\includegraphics[width=7cm, clip=true]{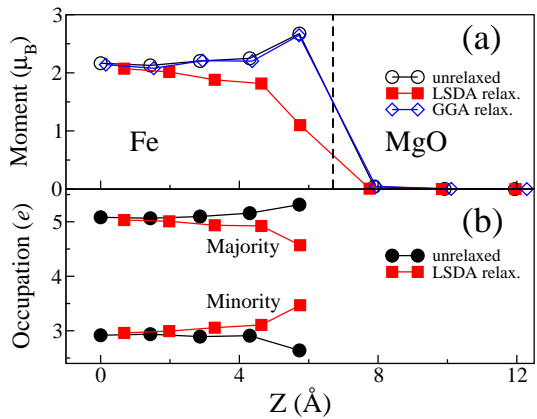}
\caption{(Color online) The magnetic moment (a) and spin-decomposed electron occupation (b) for 
a [Fe]$_{10}$/[MgO]$_6$ 
superlattice. These are calculated using Bader analysis for the valence electrons. 
The vertical dashed line in the top panel denotes the Fe/MgO interface.}
\label{moment}
\end{figure}

The local magnetic moments of the Fe atoms [Fig.~\ref{moment}(a)] are
calculated by integrating the spin densities in the domains determined by
charge densities resulting from the Bader analysis. These are similar for the
GGA-relaxed and unrelaxed structures, both presenting a significant enhancement
of the Fe magnetic moment at the MgO interface.  For the LSDA-relaxed structure
however such interfacial magnetic moment is dramatically suppressed. This is
only 1.10~$\mu_B$, to be compared with 2.65 $\mu_B$ and 2.68 $\mu_B$
respectively for the GGA-relaxed and unrelaxed geometries. Since $\Delta{\cal
Q}$ is much smaller than the change in the magnetic moment, the dramatic
decrease of interface magnetic moment in the LSDA calculations is caused by an
internal electron redistribution between the majority and minority spin
sub-bands. This is demonstrated by the spin-decomposed electron occupation
[Fig.~\ref{moment}(b)]. Since in the LSDA relaxation $d_{-1}$ is significantly
shorter than that for GGA relaxation, the suppression of the magnetic moment is
expected, based on the fact that many magnetic materials undergo magnetic
collapse under pressure.\cite{Kunes08} 

%

%
\begin{figure}[ht]
\includegraphics[width=7cm, clip=true]{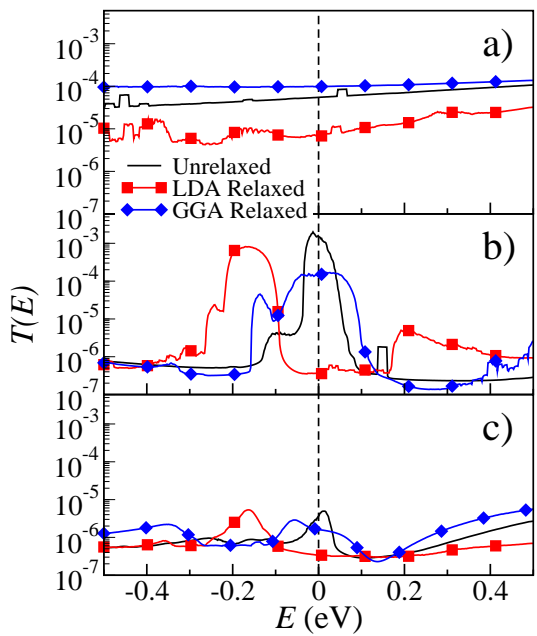}
\caption{(Color on line) Transmission $T(E)$ for PA majority spins (a), PA 
minority spins (b), and for the AA (c) 
of a Fe/MgO/Fe MTJ. The Fermi energy is set to zero.}
\label{trc}
\end{figure}

The effects of the interface structure on the electronic transport are studied
with {\sc smeagol}.\cite{Rocha06} At zero bias the conductance, $G$, is simply
$G={e^2}/{h}\:T(E_\mathrm{F})$, where $e$ is the electron charge, $h$ the
Planck's constant and $T(E_\mathrm{F})$ the transmission coefficient calculated
at $E_\mathrm{F}$.  $T(E_\mathrm{F})$ is presented in Fig.~\ref{trc}. The
transmission for majority spins in the PA depends weakly on the energy but it
is reduced by about one order of magnitude for the LSDA-relaxed structure, when
compared to both the GGA and the unrelaxed ones.  In contrast the minority spin
channel for PA is dominated by a peak at around $E_\mathrm{F}$. This originates
from IRSs at the Fe/MgO interface.\cite{Rungger07} For the LSDA-relaxed
structure these peaks are shifted 0.2~eV to lower energies with respect to the
GGA-relaxed case. A similar shift is found for both spins in the AA
configuration.  This
shift leads to a large change in the low bias conductance. As is shown in
Tab.~\ref{table2}, the total conductance of the LSDA-relaxed structure is about
30 times smaller than that of GGA-relaxed structure for the PA. Although the
magnetic moment and charge transfer are similar for the GGA-relaxed structure
and the unrelaxed structure, the transmission coefficients for the two
structures are significantly different. The GGA-relaxed structure shows less
pronounced and wider peaks in the transmission coefficient than those of the
unrelaxed structure. This results in the conductance of GGA-relaxed structure
being 5 times smaller than that of the unrelaxed one. 

\begin{figure}[ht]
\includegraphics[width=7.5cm, clip=true]{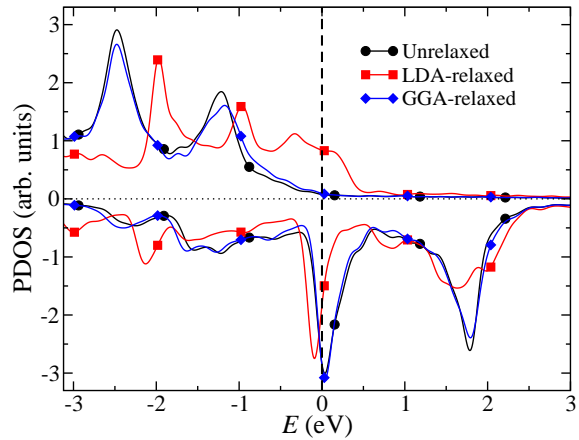}
\caption{(Color on line) The projected density of states of the interface Fe 3d orbitals 
in a [Fe]$_{10}$/[MgO]$_6$
junction. Positive and negative projected DOS represent the majority and minority components, 
respectively.}
\label{pdos}
\end{figure}
Since the peak in transmission at around $E_\mathrm{F}$ originates from the
IRSs at the Fe/MgO interface, the shift in energy of the transmission
coefficient for the LSDA-relaxed structure relative to that of the GGA-relaxed
one is the result of a high sensitivity of IRSs to the interface geometry. In
Fig.~\ref{pdos} the density of states (DOS), projected on the interface Fe
layer, is shown for both spins. For the LSDA-relaxed structure this is
shifted by about 0.2 eV with respect to that of the GGA-relaxed structure,
which causes the corresponding energy shift of $T(E)$. Since the unrelaxed and
GGA-relaxed structures are close to each other, the general features of the
projected DOS for the two are also similar.
\begin{figure}[ht]
\includegraphics[width=6cm, clip=true]{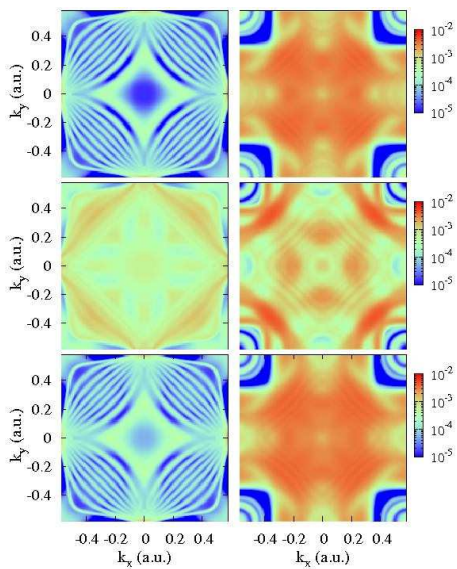}
\caption{
(Color on line) The ${\mathbf k}$-resolved projected density of states for the
interface Fe ML in a Fe/MgO/Fe junction. Unrelaxed, LSDA-relaxed and
GGA-relaxed structures are shown in the top, middle and bottom panels
respectively. The left and right panels are for majority and minority spins,
respectively.}
\label{kpdos}
\end{figure}
In order to get a better insight into the role of relaxation, in
Fig.~\ref{kpdos} we show the ${\mathbf k}$-resolved projected DOS at
$E_\mathrm{F}$ for the interfacial Fe layer. The results for the unrelaxed and
GGA-relaxed structures are rather similar, but  they differ remarkably from
that for the LSDA-relaxed case. In general the LSDA-relaxed geometry shows a
substantial reduction of the spin-polarization, demonstrated by the fact that
the ${\mathbf k}$-resolved DOS is similar for the majority and minority spins.
This is expected to produce a reduction in spin-filtering and as a consequence
a reduction in TMR. It is also important to note that the main differences
between LSDA-relaxed and GGA-relaxed (and unrelaxed) geometries are more
evident around Brillouin zone center, i.e. for electrons with a large linear
momentum perpendicular to the MgO barrier and therefore with a larger
transmission probability.

The changes in conductance lead to very different values for the TMR. Our
calculated values for the TMR are shown in Tab.~\ref{table2}. The LSDA-relaxed
structure has a TMR of about 10$^3$, which is much smaller than the TMRs of the
other two structures, and in much closer agreement to
experiments.\cite{Yuasa04} The dramatic reduction of TMR for the LDA-relaxed
structure originates from two features: the shift to lower energies of the IRSs
transmission peak, and the large reduction of the majority spin transmission
for PA.  These can be both associated with a loss of spin-polarization of the
first Fe layer at the Fe/MgO interface in the LSDA-relaxed case.
\begin{table}
\caption{
The TMR and zero bias conductance for PA and AA a Fe/MgO/Fe MTJ. GGA PBE is
used for all transport calculations. The results are shown for the unrelaxed,
the LSDA-relaxed and for the GGA-relaxed structures. The experimental TMR is
for a Fe/MgO/Fe MTJ with a 2.3 nm thick MgO barrier and a temperature of
20~K.\cite{Yuasa04} 
}
\label{table2}
\begin{center}
\begin{tabular}{cccr}
\hline\hline
         & $G_\mathrm{PA}$                & $G_\mathrm{AA}$              & $\mathrm{TMR}$(\%) \\
\hline
unrelaxed   & $1.21\times 10^{-3}$ & $7.18\times 10^{-6}$ & 17,300  \\ 
LSDA relax.      & $7.32\times 10^{-6}$ & $6.74\times 10^{-7}$ & 986 \\ 
GGA relax.       & $2.38\times 10^{-4}$ & $3.23\times 10^{-6}$ & 7,270 \\ 
Expt.            & -                    &   -             & 247 \\ 
\hline\hline
\end{tabular}
\end{center}
\end{table}

In conclusions, motivated by the significant difference between theory and
experiments on the reported TMR values in Fe/MgO/Fe(100) MTJs, we studied the
interface structure and its effects on the electronic and  transport properties
by means of DFT and the NEGF formalism. In general LSDA- and GGA-relaxed
interfacial geometries are rather different, yielding different charge transfer
and interfacial magnetism. In particular the local magnetic moment of the
interfacial Fe ML is severely suppressed for the LSDA-relaxed structure and
enhanced for GGA-relaxed ones. These differences are reflected in the transport
properties. In particular the differences at the interface determine the energy
position of resonances in the transmission for the minority spins. These
resonances are caused by IRSs and largely determine the zero-bias conductance.
As a consequence the TMR is rather sensitive to the interfacial structure.
These features can partially explain the disagreement between theory and
experiments  and are expected to apply to many other systems relevant for 
spintronics

X.F, O. B, M.A., and S.L acknowledge an ANR grant  ANR-06-NANO-053, and I.R and
S.S thank Science Foundation of Ireland for financial support (grant nr.
07/IN.1/I945 and 07/RFP/PHYF235).


\end{document}